\documentstyle[aaspp4,aabib]{article}
\input epsf

\def\s2n{S^{\prime}/N}
\def\vecr{{\bf r}}
\def\vecdr{{\bf \Delta r}}
\def\dr{{\Delta r}}
\def\aq{{\langle Q \rangle}}

\begin{document}
\title{The Effects of Noise and Sampling\\ on the Spectral Correlation
Function}

\author{Paolo Padoan\footnote{ppadoan@cfa.harvard.edu},
Erik W. Rosolowsky\footnote{eros1@uclink.berkeley.edu}$^,$\footnote{Current
Address: UC
Berkeley Astronomy Department, 601 Campbell Hall, Berkeley, CA 94720}
and Alyssa A.Goodman\footnote{agoodman@cfa.harvard.edu}$^,$\footnote{National
Science Foundation Young Investigator}}
\affil{Harvard-Smithsonian Center for Astrophysics, Cambridge, MA 02138}

\begin{abstract}

The effects of noise and sampling on the ``Spectral Correlation Function''
(SCF) introduced by Rosolowsky et al. 1999 are studied using observational 
data, numerical simulations of magneto--hydrodynamic turbulence, and simple 
models of Gaussian spectral line profiles. The most significant innovations 
of this paper are: i) the normalization of the SCF based on an analytic 
model for the effect of noise; ii) the computation of the SCF as a function 
of the spatial lag between spectra within a map. 

A new definition of the ``quality'' of a spectrum, $Q$, is introduced, 
which is correlated with the usual definition of signal--to--noise.
The pre--normalization value of the SCF is a function of $Q$. 
We derive analytically the effect of noise on the SCF, 
and then {\it normalize} the SCF to its analytic approximation.

By computing of the dependence of the SCF on the spatial lag, 
$S_0(\dr)$, we have been able to conclude that:

\begin{itemize}
\item $S_0(\dr)$ is a power law, with slope $\alpha$, in the range of 
scales $\ell_i<\ell<\ell_o$.
\item The correlation outer scale, $\ell_o$, is determined by the size
of the map, and no evidence for a true departure from
self--similarity on large scales has been found.
\item The correlation inner scale, $\ell_i$, is a true estimate of the
smallest self--similar scale in a map.
\item The spectral slope, $\alpha$, in a given region, is independent of velocity
resolution (above a minimum resolution threshold), spatial resolution, 
and average spectrum quality.
\item Molecular transitions which trace higher gas density yield larger
values of $\alpha$ (steeper slopes) than transitions tracing lower gas density.
\item Nyquist sampling, bad pixels in detector arrays, and reference 
sharing data acquisition need to be taken into account for a correct 
determination of the SCF at $\dr =1$. The value of $\alpha$, however,
can be computed correctly without a detailed knowledge of observational
procedures.
\end{itemize}

\end{abstract}


\section{Introduction}

Rosolowsky et al. (1999) (RGWW) have recently introduced a new method to
analyze large maps of molecular spectral lines. They propose to use the
``Spectral Correlation Function'' (SCF) as a way to test theoretical
models against observational data. The SCF at a given position in a
map is defined as the quadratic sum of the difference between the
spectral line profile at that position and the profile at neighboring
positions:
\begin{equation}
S_0(T_1,T_0)=1- \sqrt{\frac{\Sigma_v(T_1(v)-T_0(v))^2}{\Sigma_v
T_1(v)^2+\Sigma_v
      T_0(v)^2}}
\label{for1}
\end{equation}
where $T_i(v)$ is the antenna temperature at the velocity channel $v$ at the
position $i=0,1$ in the map. The definition of the SCF in RGWW
is more general than this, since it allows for translation along the velocity
axes and rescaling of antenna temperature that minimize the difference
between neighboring spectra. In the present work we only discuss
the SCF as defined in equation (\ref{for1}), corresponding to
``$S^0$'' in RGWW. 

The SCF is similar in spirit to some of the analysis tools used to
extract clumps from 3-D spectral line data cubes (Stutzki \& Gusten 1990;
Williams, De~Geus \& Blitz 1994), in that it makes direct use 
(no transform involved) of both spatial and velocity information. 
However, the SCF is different from 
these methods, because it makes no attempt to extract the properties of 
clumps from data cubes; it simply compares
neighboring spectra with each other, utilizing both the spatial and
velocity dimensions. 

Some previous statistical analyses do not explicitly 
use the velocity dimension in analyzing spectral line cubes 
(for example the wavelet analysis by Gill \& Henriksen 1990 and 
Langer, Wilson, \& Anderson 1993; the structure-tree statistics by 
Houlahan \& Scalo 1992; the column density distribution by 
Blitz \& Williams 1997; the $\Delta$--variance method by Stutzki et al. 1998
and Mac Low \& Ossenkopf 1999; the fractal analysis by Beech 1987, 
Bazell \& D\'{e}sert 1988, Scalo 1990, Dickman, Horvath \& Margulis 1990,
Falgarone, Phillips \& Walker 1991, Zimmermann, Stutzki \& Winnewisser 1992, 
Henriksen 1991, Hetem \& Lepine 1993, Vogelaar \& Wakker 1994, 
Elmegreen \& Falgarone 1996). 

Other statistical analysis make 
use of the velocity dimension, but only to estimate the centroid velocity of 
the spectra, and compute the structure (or autocorrelation) function of 
velocity fluctuations (Scalo 1984; Kleiner \& Dickman 1985, 1987; 
Hobson 1992; Miesch \& Bally 1994), or the distribution of line 
centroids (Miesch \& Scalo 1995; Lis et al. 1996; Miesch, Scalo \& Bally 1999).
The moment analysis computes statistical moments
(velocity centroids, line width, skewness and kurtosis) of single spectra
in a map, and derive their distribution over the whole map 
(Falgarone et al. 1994; Padoan et al. 1999), independent of their position
in a map.

A method that exploits both velocity and spatial information is the
Principal Components Analysis by Heyer \& Schloerb (1997). This method 
describes clouds as a sum of orthogonal functions in a manner mathematically 
similar to wavelet analysis, and is the most promising way to extract the 
power spectrum of molecular cloud turbulence from observational data. 

Therefore, prior to the SCF, no statistical analysis had  
tackled the problem of quantifying spatial correlations in 
spectral maps, taking into account also the velocity information, with the
exception of the Principal Component Analysis. The SCF
quantifies the correlation between spectra at a given distance from each other
(spatial lag), using the full velocity profile information, because the
comparison between spectra is made channel to channel.

\nocite{Gill+Henriksen90}
\nocite{Langer+93} \nocite{Houlahan+Scalo92} \nocite{Blitz+Williams97} \nocite{Stutzki+98}
\nocite{MacLow+Ossenkopf99} \nocite{Beech87}  \nocite{Bazell+Desert88}  \nocite{Scalo90}
\nocite{Dickman+90}  \nocite{Falgarone+91} \nocite{Zimmermann+92} \nocite{Henriksen91}
\nocite{Hetem+Lepine93} \nocite{Vogelaar+Wakker94} \nocite{Elmegreen+Falgarone96}
\nocite{Scalo84} \nocite{Kleiner+Dickman85} \nocite{Kleiner+Dickman87}
\nocite{Hobson92} \nocite{Miesch+Bally94} \nocite{Heyer+Schloerb97}
\nocite{Falgarone+94} \nocite{Padoan+98per}

RGWW concluded that the SCF can find differences between observational data 
and synthetic spectra from numerical simulations of turbulent flows that are 
not found by other statistical analyses of line profiles (e.g. the moment 
analysis by Falgarone et al. 1994). That preliminary result motivates the 
present work, where we try to study in more detail 
the effect of noise and sampling on the SCF, and to improve on the first 
implementation of the method. The main issue is that the value of the SCF 
as defined in RGWW depends on the signal to noise ($S/N$), as illustrated 
in Figure~1 of both that paper and Figure~\ref{fig00} of this paper. A 
simple way to eliminate this signal-to-noise dependence, proposed in RGWW, 
is to make the $S/N$ uniform over the map, by adding noise. In RGWW, spectra 
with $S/N<5$ were discarded, and noise was added to higher quality
spectra to force $S/N=5$. In the present work, instead, we have analytically
estimated the main effect of noise on the SCF, using a novel definition
of the $S/N$. We can therefore compute a noise--corrected SCF that is hardly
dependent on the $S/N$, without adding any extra noise to the data.

In the new-and-improved implementation of the SCF offered in this paper,
when observational data are compared with synthetic maps, noise must be
added to the synthetic data, under the assumption that it is uniform
over the observed map. If noise is not uniform over the observed map, 
or if it is correlated over a few map positions, extra noise must be added 
to the observational spectra until the noise is both spatially uniform
and uncorrelated. However, we do not need to adjust the $S/N$ to be
uniform as in RGWW.

In the following two sections, the effect of noise on the SCF is
discussed, and an analytic model is computed that allows
the values of the SCF to be corrected for the effect of noise.
In \S 4 we study the effects of both velocity sampling and 
spatial lags between spectra. We show that a correlation 
inner scale, $l_i$, can be defined. On scales smaller than $l_i$
the spectral map is not self-similar. Results are tested in \S 5, 
by computing the SCF using two maps of the Rosette Molecular Cloud
with different resolution. A discussion is presented in \S 6, and
in \S 7 we summarize our conclusions.

\section{Effect of Noise}

In RGWW the signal--to--noise of a spectrum is defined in a 
conventional way, computing the signal
as the maximum antenna temperature of a Gaussian fit to each line profile,
while in the present work we define the signal--to--noise in a different way.
We characterize the signal--to--noise by spectrum quality, $Q$, computed
as the rms of the antenna temperature for all channels inside a velocity
window, divided by the rms noise over the whole map. This definition
will prove very useful in the analytic computation presented in \S 3,
where its relation with the usual definition of signal--to --noise is 
discussed further.

The value of $T_i(v)$ in eq. (\ref{for1}) is the sum of signal and
noise, and therefore two intrinsically identical spectra can have significant
channel to channel differences due to the noise alone.
The noise has a stronger effect
for spectra with low $S/N$ than for spectra with high $S/N$, which is why
the SCF increases gradually with the $S/N$, as illustrated in Figure~1 of RGWW.
The top panel of Figure~1 of the present paper, which shows $S_0$ as a function
of spectrum quality $Q$, is very similar to the left panel of RGWW
Figure~1. The only difference is in how spectrum quality is defined:
$S/N$ in RGWW and $Q$ here.

In Figure~\ref{fig00} the SCF is plotted versus
$Q$, for the C$^{18}$O (1-0) spectra in a map of Heiles Cloud 2
(deVries et al. 2000).\nocite{deVries+99} The upper panel shows the SCF
versus $Q$ for
data in their original (real) positions, and the lower panel for
a ``randomized'' map created by randomizing the positions of the spectra.
These randomized maps proved useful comparison tools in RGWW. Statistics
which only consider distributions of line parameters (such as
moment analyses, Falgarone et al. 1994; Padoan et al. 1998) 
would find the original and randomized maps to be identical,
while the SCF can, and always does, find them different.

A map of ``synthetic spectra'' can be made as a collection of
Gaussian profiles with a linear gradient in their amplitude
along one spatial direction.
The resulting map is very smooth. The SCF of such an artificial
map is very close to 1 ($\approx 0.99$) everywhere on the map,
assuming the gradient is small enough.
Then noise can be added to the spectra, and suddenly the SCF
changes dramatically: it decreases towards zero for very low
$Q$, and its values are scattered around an average value for
any given $Q$. This is qualitatively very similar to the
dependence of the SCF on $Q$ computed for observational data.
Does this mean that the SCF versus
$S/N$ is only telling us about instrumental noise, and not about
the physics of molecular clouds? The answer is no, and this is
illustrated in Figure~\ref{fig6}.

The solid line in the left panel of
Figure~\ref{fig6} shows the SCF versus the rms antenna temperature
(there is no noise in the artificial spectra) for each spectrum in
the smooth map of Gaussian spectra (see Figure~\ref{fig6} captions
for details). The scatter plot in the same panel shows instead
the SCF versus the rms antenna temperature (still no noise added)
for each spectrum of a more realistic map of $^{13}$CO (J=1--0)
synthetic spectra that are
computed using the results of MHD simulations of super--sonic
turbulence\footnote{These synthetic spectra are calculated from data cubes
obtained as results of super--sonic MHD simulations with sonic rms 
Mach number equal to 10.6, and assuming an average gas density equal 
to $540$~cm$^{-3}$, and a physical size of the
simulated box of 3.7 pc, without including self--gravity, stellar 
radiation, or stellar outflows. The original numerical mesh is 128$^3$
in size, while the maps contain 90$\times$90 spectra (the numerical mesh
has been rebinned to 90$^3$ for the computation of the radiative transfer).
Details of the computation of synthetic spectra using simulations of 
MHD turbulence can be found in Padoan \& Nordlund 1999; Padoan et al. 1999.}
(Padoan et al. 1998).
The synthetic spectra from MHD simulations are far from being perfectly
smooth Gaussians. They have features
created by the combination of the projected density field
and the radial velocity distribution along the line of sight,
such as non Gaussian spectral wings and
multiple components. Neighboring spectra in the MHD simulations
can differ because their shapes are different, their
integrated temperatures are different, or their centroid velocities
are shifted. These are the reasons why the SCF versus antenna temperature
in the realistic synthetic spectra has a large scatter and a smaller
average value than the SCF of the Gaussian spectra that is equal
to 0.99 everywhere on the map. The difference between the two cases
can still be appreciated after the noise is added to the spectra
(right and central panels of Figure~\ref{fig6}). So, although the noise
is responsible for the gross ``rising'' dependence of the SCF on $Q$,
the actual distribution of the SCF at a given $Q$ depends also on the
type of structures present in the spectral map, and thus on
the physics that generates spectra with those particular structures.

\section{How to Correct for the Effect of Noise}

In order to quantitatively study the intrinsic differences between
distributions of the SCF for different observational spectral maps,
it is useful to define a new SCF, corrected
for the effect of noise. One way to achieve this is to provide
a simple analytic model of the effect of noise.

First consider the definition of the SCF (eq. \ref{for1}).
\begin{eqnarray}
S_0(T_1,T_0)&=&1-\sqrt{\frac{\sum_v (T_1-T_0)^2}{\sum_v T_1^2+\sum_v
T_0^2}}\\
&=&1-\sqrt{1-\frac{2\sum_v T_1\cdot T_0}{\sum_v T_1^2+\sum_v T_0^2}}
\end{eqnarray}
In order to simplify the expression further, a given spectrum,
$T_i(v)$, can be theoretically separated into a signal component ($s$) and
a noise component ($\mu$): $T_i(v)=s_i(v)+\mu_i(v)$.\footnote{$v$ is a
discrete variable; we write it as an argument to be consistent with common
notation in the literature. The subscript $i$ refers to the position
of the spectrum in the map.}
For purposes of this derivation, we assume the noise function to have a
mean value of zero and an rms value (over the whole map) equal to $N$. 
({\it Throughout this paper, the letter ``$N$" is used to mean rms 
``noise": it has nothing to do with a number of channels or samples.})  
In addition, we assume that the noise in any spectrum is
uncorrelated with the noise in  another spectrum.  Using our assumptions
about the noise,  we can set
$\sum_v  s_i
\mu_i=0$  for any $s_i$ that varies slowly in velocity space because
the mean value of $\mu_i$ over any interval is zero.  Using the definition
of $N$
as rms baseline noise, we see that $\sum_v
\mu_i^2dv= WN^2$, where $W$ is the velocity range over
which the SCF is evaluated, and $dv$ the width of the velocity
channels. (The same value of $W$ is used for all
spectra).  Finally, we can approximate $\sum_v
\mu_i\mu_j =0$ because the two noise functions are uncorrelated
and hence their product will be normally distributed around zero.

With all of these simplifications, the definition of the SCF can be
reduced to the following:
\begin{eqnarray}
S_0(T_1,T_0)&= & 1-\sqrt{1-\frac{2(\sum_v s_1 s_0 + \sum_v s_1
\mu_0 + \sum_v s_0 \mu_1 +\sum_v \mu_1\mu_0)}{\sum_v T_1^2+\sum_v
T_0^2}}\\
&\approx& 1-\sqrt{1-\frac{2\sum_v s_1 s_0}{\sum_v T_1^2+\sum_v T_0^2 }}
\end{eqnarray}

In this paper, we have chosen to modify the definition of
signal-to-noise (see \S 2). We refer to the new $S/N$ as ``spectrum
quality'', $Q$:
\begin{equation}
Q_i
=\frac{1}{N}\sqrt{\frac{\sum_v T_i(v)^2dv}{W}}
\qquad \Longrightarrow \qquad \sum_v T_i(v)^2dv =
WN^2Q^2_i
\end{equation}
The spectrum quality, $Q$, is compared with the traditional $S/N$ 
based on Gaussian fits to the spectra (eg RGWW) in Figure~\ref{fig10}, 
using the spectra from the map of Heiles Cloud 2. Figure~\ref{fig10} 
shows that $Q$ depends on the choice of the velocity window, decreasing 
as $W$ increases, and it is in general smaller than the usual definition 
of $S/N$. In this work, we use $W=10\sigma$, where $\sigma$ is the standard 
deviation in velocity of a spectrum created by averaging
over the whole map ($W\approx 6\sigma$ was used in RGWW).

We can expand $\sum_v T_i(v)^2$ and use our assumptions about
the noise to simplify our results.
\begin{equation}
\sum_v T_i(v)^2=\sum_v \left[s_i(v)+\mu_i(v)\right]^2=\sum_v s_i^2 +\sum_v
2s_i\mu_i+\sum_v \mu_i^2=\sum_v s_i^2 + WN^2/dv
\end{equation}
Combining the results from equations 6 and 7 yields the following:
\begin{equation}
\sum_v s_i^2= {WN^2\over{dv}}\left[Q^2_i-1\right]
\end{equation}
In order to normalize the SCF, we want to quantify the effect of noise alone, 
and not the effect of intrinsic variations between different spectra. We therefore 
consider the case of neighboring spectra that are identical in their signal
(they differ only for the noise component), 
$\sum_v s_1 s_0 = \sum_v s_1^2 = \sum_v s_0^2$. In that case, the SCF is: 
\begin{eqnarray}
S_{0,max}(Q) & = &
1-\sqrt{1-\frac{ 2 WN^2\left[Q^2-1\right]}
{2WN^2Q^2}} \\
& = & 1-\frac{1}{Q}
\end{eqnarray}

Notice that we only use the rms value of the noise averaged over the whole 
map, $N$, and not the specific noise level at the position of each 
spectrum. Observational data usually contain non--uniform noise,
that is the noise level of two different spectra can be slightly different.
However, spatial fluctuations of noise on spectral maps of good
quality are usually small (about 10\% in the case of the Heiles Cloud
2 used in this work), and the effect on the SCF method are 
very small (we have verified this in a number of experiments, by
normalizing the SCF with the value of the local noise at each 
spectrum position, and by adding spatially non--uniform noise
to maps of synthetic spectra). We therefore use only the global
rms noise value $N$, which allows the derivation of the analytic
expression (10).  
The function in eq. (10) should maximize the SCF, because its derivation
assumes that neighboring spectra are identical, in the absence of noise.
In fact, the dependence of the SCF on $Q$ is perfectly fit by this simple 
function, for the smooth Gaussian model (see Figure~\ref{fig2}). 
Given the assumption of  identical
neighboring spectra, all observational data cubes and realistic models
are expected to yield values of the SCF that are not larger than the one
given by the simple function here derived. However, the sums
involving the noise (the $\mu$ terms) that have been eliminated because
they are approximately zero for uncorrelated noise, can have both positive and
negative deviations from zero, not taken into account in the
present derivation. Such random deviations explain the scatter in the
plot of Figure~4, and they are the reason why the SCF of observational
data can be even larger than its value predicted analytically, 
for very low values of $Q$ (left panels of Figure~\ref{fig9}).

Since the simple function in equation (10) explains most of the effect of
noise on the SCF, it can be used as a reference, and a ``noise-corrected"
SCF can be expressed in terms of deviations from that reference function. 
In practice we divide the value of the ``raw'' SCF at any given $Q$, by its 
expected value according to equation (10) to
derive the {\it normalized} SCF. Normalized SCF distributions 
are shown in Figure~\ref{fig9} for  the MHD model and for Heiles Cloud 2. 
The Heiles Cloud 2 map
has been decreased in resolution by a factor of two, to eliminate
spatial correlation of noise, as discussed
in \S 6, and uniformly distributed Gaussian noise has been added
to the spectra, with an rms value equal to the rms noise of the
original data, to completely eliminate any possible residual
correlation in the noise. As a result, the values of $Q$ span a range
from 1 to about 6. Spatially uniform Gaussian noise has been added
also to the numerical MHD and Gaussian models shown in Figure~5, 
in order to obtain a range of values of $Q$ for
the synthetic spectra between 1 and 6, as in the observational ones.

In the rest of this paper, average values of the SCF are computed using 
only spectra with $Q>2$, since for smaller values of $Q$ the SCF is 
dominated by noise. Figure~6 shows that the value of the normalized SCF,
averaged over the whole map, is roughly independent of $Q$ (at least for $Q>2$), 
and therefore the normalization based on the analytic formula (10) is able 
to eliminate the gross dependence of the SCF on the noise. 
The plot is computed using a map of 90x90 synthetic $^{13}$CO spectra, 
from Padoan et al. (1999). The value of $\aq$ is varied by
adding different levels of noise to the synthetic spectra.

\section{Effect of Sampling}

\subsection{Velocity Window}

In RGWW the signal--to--noise ($S/N$) is defined by computing the signal as
the maximum antenna temperature of a Gaussian fit to each spectrum, and the
SCF is computed using only velocity channels within 3 FWHMs of the velocity
centroid, where both the velocity centroid and the FWHM are results of a Gaussian
fit to each spectrum. In the present work no Gaussian fits are used,
and the signal--to--noise is characterized by spectrum quality, $Q$,
defined in eq. (6). $Q_i$ has been computed as the rms of antenna
temperature for all channels inside a velocity window, $W$, at the map
position $i$, divided by the rms noise over the whole map, $N$.

Our definition of $Q$ takes into account the fact that all
velocity channels inside the velocity window are used in the comparison
of two neighboring spectra, and so the spectrum ``quality'' 
depends in part on the width of the velocity window.
The velocity window used in the computation of the SCF has the same width, $W$,
for all spectra in the map, and it is centered around the velocity centroid of
each map position. As a result, one advantage of using $Q$ instead of 
the usual $S/N$, is that the SCF
becomes independent of the choice of the value of $W$ (except in the
case of the randomized maps). Moreover,
this definition of $Q$ allows the analytic computation of the effect
of noise on the SCF, as shown in the previous section.
The dependence of the SCF versus $S/N$ on $W$
is due to the fact that a larger velocity window
introduces more velocity channels that are dominated by noise than a smaller
velocity window, and therefore decreases the value of the SCF. This does not
occur when $Q$ is used instead of the usual $S/N$, because the value of
$Q$ decreases as $W$ increases.

In Figure~7 the average value of the normalized SCF for the Heiles
Cloud 2 C$^{18}$O map is plotted versus the value of $W$. Spectra
with $Q<2$ are not used because their SCF is too strongly affected by the
noise. The SCF averaged over the whole map, is not affected 
by the value of $W$. (``Error bars'' in Figure~7 show the 1--$\sigma$ 
dispersion in the SCF for each value of $W$). On the other hand, the 
SCF of the spectra with randomized positions decreases with 
increasing $W$. This is mainly due to the fact 
that, once the position of the spectra are randomized, neighboring 
spectra can have values of $Q$ very different from the 
one of the reference spectrum. We have verified that if $Q$ is defined as
the average of the quality of all spectra used to compute the SCF at any 
position in a map, than the SCF is roughly independent of $W$ also 
for the case of spectra with randomized positions.

Although the SCF varies extremely little in the range $1<W/\sigma<30$, 
it is better to use the same value of $W/\sigma$ when comparing 
observational data
with theoretical models, if the values of the SCF for the randomized
spectra are to be compared. Note that in RGWW's analysis of the Heiles
Cloud 2 map, the value of $W$
is essentially constant, because variations in the FWHM
of different spectra are rather small.

\subsection{Spatial Lags}

In the previous sections we have shown that the value of the normalized
SCF does not 
depend strongly on the average signal--to--noise (or spectrum quality) of 
the data or on the width of the velocity window used to compare the spectra. 
We have used values of the SCF averaged over the whole spectral map,
and only adjacent spectra have been compared. The physical separation
between adjacent spectra (or pixels in a map) depends on the distance
to the observed cloud, on the size of the telescope beam, and on the 
way the cloud has been sampled in the observations. A different
physical separation between spectra yields different values of the SCF.
In this section, we compute the SCF using different values 
of the distance between spectra in a map, that we call the ``lag'', or
$\Delta r$ ($\Delta r=1$, in pixel units, in the previous sections),
and we explore how the SCF depends on lag. The value of the SCF at
a position $\vecr$ in a map and lag $\dr$ is given by the expression:
\begin{equation}
S_0(\vecr,\dr)=\left\langle 1- \sqrt{\frac{\Sigma_v[T(\vecr,v)-T(\vecr+\vecdr,v)]^2}
{\Sigma_vT(\vecr,v)^2+\Sigma_vT(\vecr+\vecdr,v)^2}} \right\rangle _{\vecdr}
\label{}
\end{equation}
where the average is done for all vectors $\vecdr$ of length $\dr$. The
value of the SCF for lag $\dr$, averaged over the map is:
\begin{equation}
S_0(\dr)=\left\langle S_0(\vecr,\dr) \right\rangle _{\vecr}
\label{}
\end{equation}

Figure~8
shows the SCF versus $\Delta r$ for the spectral map of the Heiles Cloud 2,
the MHD model, and the smooth Gaussian model. The Gaussian model has no 
spatial structure, apart from a smooth one--dimensional intensity gradient,
and therefore yields values of the SCF that are hardly dependent on $\Delta r$.
The lag dependence is instead stronger in the real data and in the MHD model,
since both contain strong spatial and velocity structures. For both the 
C$^{18}$O data and the MHD model, $S_0(\dr)$
can be well approximated by a power law inside the range of scales
$\ell_i<\ell<\ell_o$, where we call $\ell_i$ the correlation
inner scale, and $\ell_o$ the correlation outer scale. The C$^{18}$O 
map of the Heiles Cloud 2 has a SCF--lag power law slope of 
$\alpha\approx -0.17$, and flattens at $\ell_o \approx 11$, 
in pixel units, that corresponds to $\ell_o \approx 0.4$~pc. 
The slope and the correlation scale for the MHD model are similar, 
but should not be compared directly to the values found for the 
Heiles Cloud 2, since they are computed for a different molecular 
transition, $^{13}$CO. The MHD model is here plotted as an illustration, 
and detailed comparisons between models and observations 
will be included in our next paper.

$S_0(\dr)$ depends on the particular molecular transition used to map a cloud. 
At fixed angular resolution, molecular transitions which probe preferentially 
regions of high gas density generate maps with ``sharper'' structures, 
and larger values of the spectral slope $\alpha$, than transitions which probe
low gas density. In Figure~9, maps of synthetic $^{12}$CO, $^{13}$CO, 
and CS spectra, from Padoan et al. (1999), have been used to compute 
$S_0(\dr)$. The same three dimensional cloud model, obtained as the result of
numerical simulations of super--sonic MHD turbulence (Padoan et al. 1998),
has been used in all three cases. Figure~9 shows that synthetic
spectra computed for a particular molecular transitions should be 
compared only with observational spectra of the same molecular transition, 
or a {\it very} close substitute.

We have seen in \S 6 (Figure~3) that the value of the SCF for $\dr =1$
is roughly independent on the value of the spectrum quality averaged 
over the whole map, $\aq$. The same is true also for the spectral 
slope $\alpha$. In Figure~10 we have plotted $S_0(\dr)$ for the synthetic
map with different levels of noise. Noise is added to the synthetic
spectra, in order to obtain different values of the average spectrum 
quality: $\aq=1.5,\, 3.0,\, 6.0,\, 12.0$. The value of $\alpha$ 
is almost constant. It tends to slightly decrease with decreasing
$\aq$, but it grows again for $\aq<3.0$. The values of $S_0(\dr)$
increase with decreasing $\aq$, but they tend to stabilize
around $\aq=3.0$. Typical values of the average spectrum quality
in observational data are around $\aq=3.0$, and we find that both
$S_0(\dr)$ and $\alpha$ are typically not effected very much by
variations of $\aq$.

In synthetic maps of $90\times 90$ spectra, or in observational
maps with a comparable number of spectra, the power law shape of
$S_0(\dr)$ spans about an order of magnitude in scale, and it would
probably extend to a scale much larger than $\ell_o$, if the 
spectral map covered a larger region. In all maps analyzed so far,
we have found that $\ell_o$ is related to the size of the map,
and therefore it is not a true estimate of the largest self--similar 
scale. The correlation inner scale, $\ell_i$, instead, must be an
intrinsic smallest self--similar scale, {\it for a particular tracer}, 
rather than an artifact of the spatial resolution or of the size 
of the map. The effect of finite resolution can only be that of 
increasing the value of the SCF for any given $\dr$, relative to 
an ideal case of infinite resolution. The flattening of $S_0(\dr)$ 
for small $\dr$ is also consistent with the condition $S_0(\dr)<1$, 
which forces $S_0(\dr) \rightarrow 1$ on small scale. In the case of 
the Heiles Cloud 2 C$^{18}$O map, the smallest self--similar scale 
is $\ell_i\approx 0.06$~pc. This interpretation of $\ell_i$ and 
$\ell_o$ is further confirmed in \S 5.  

Since $S_0(\dr)$ can be fitted by a power law, with the exponent 
roughly independent of $\dr$ for a range of scales, $\ell_i<\ell<\ell_o$,
the correlation properties of a spectral map can be quantified by the
SCF slope $\alpha$, independent of the spatial resolution of the map 
(or telescope beam), or the exact distance to the observed cloud. 
Although the pixel size should be taken into account when comparing 
data and theoretical models, or different clouds, the uncertainty 
in the distance to the observed cloud should have almost no effect
on the determination of the spectral slope $\alpha$.

\subsection{Spatial and Velocity Resolution}

In order to verify that the spectral slope $\alpha$ does not vary
significantly with the spatial resolution of a map, we have 
computed $S_0(\dr)$ for the Heiles Cloud 2 map and the
synthetic map first at full resolution, and then at a resolution
three times worse than the original one, by rebinning the maps 
into a smaller number of spectra (for example if the map size is
reduced by a factor of three, each spectrum of the smaller map is
computed as the average of all the 3x3 spectra around its position 
in the original map). $S_0(\dr)$ for the original maps
is represented with square symbols in Figure~11, while $S_0(\dr)$ 
for the smaller rebinned map is represented with asterisks.
A variation of a factor three in resolution has no effect on the 
spectral slope of the Heiles Cloud 2 map, $\alpha=-0.16$ (left panel), 
while the effect is very small for the synthetic map, 
from $\alpha =-0.20$ at high resolution, to $\alpha =-0.21$ at small 
resolution. Moreover, once $S_0(\dr)$ is rescaled into physical units of 
length, its actual value at any given length is unchanged, when the 
spatial resolution is changed by a factor of three, in the case of
the Heiles Cloud 2 map, and varies only by 1-2\%, in the case of the 
synthetic map. 

We have also computed $S_0(\dr)$ in both maps for different 
velocity resolutions, by rebinning the original data cubes 
into a smaller number of velocity channels. We find insignificant
variations of $S_0(\dr)$, when the number of velocity channels is
varied by a factor of 10. Results of this computation are not 
presented in a figure, since the values of $S_0(\dr)$ for velocity
resolutions differing by a factor of 10 basically overlap 
on the plots. This could be due to the fact that for a specific 
tracer, mapped with a finite spatial 
resolution, and finite signal-to-noise, there may be a limit to the 
"resolvable" velocity structure in a map, due essentially to spatial 
blending of what would otherwise be distinct velocity components.
We conclude that the spectral slope $\alpha$ is almost
independent of spatial and velocity resolution, and that the 
same is true for the actual values of $S_0(\dr)$, if rescaled 
to physical units of length.

\section{The Rosette Molecular Cloud: a Test of the SCF}

In the previous section we have shown that i) $S_0(\dr)$ is 
well approximated by a power law in the range of scales
$\ell_i<\ell<\ell_o$; ii) the correlation outer scale $\ell_o$ 
is determined by the size of the map; iii) the correlation 
inner scale $\ell_i$ is a true estimate of the smallest 
self-similar scale on the map; iv) the spectral slope $\alpha$
is almost independent of spatial resolution, velocity resolution,
and average spectrum quality $\aq$.  
In the present section we test these results, using two $^{13}$CO
(J=1-0) maps of the Rosette Molecular Clouds, obtained by Blitz \& Stark
(1986), and by Heyer et al. (2000).

The Rosette map by Heyer et al. covers a region that is approximately the
same as the one covered by the Blitz \& Stark map in galactic 
longitude, and about 4 times smaller in galactic latitude.
The spatial and velocity resolutions in the map by Heyer et al.
are higher than in the map by Blitz \& Stark: 0.23 pc and 0.06 km/s
versus 0.7 pc and 0.68 km/s respectively (assuming a distance to the 
Rosette Molecular Cloud of 1600 pc). The average spectrum quality
of the Heyer et al. map is $\aq=2.83$, and $\aq=2.89$ for the 
portion of the Blitz \& Stark map, that corresponds to the 
region mapped by Heiles et al.
The considerably different spatial and velocity resolutions of the 
two maps make them suitable for testing the SCF; moreover, the very 
similar average spectrum quality guarantees that noise does not
effect the comparison of the two maps at all.

In Figure~12, $S_0(\dr)$ is plotted for both maps. The unit of length
in the left panel is the pixel size of the map, while in the right panel
it is the physical length in parsecs, assuming a distance of 1600 pc. 
We have also plotted $S_0(\dr)$
computed on a portion of the Blitz \& Stark map, that corresponds
exactly to the area mapped by Heyer et al. (triangle symbols).
The two maps yield practically indistinguishable values of $S_0(\dr)$,
when limited to the same region, which demonstrates that the $S_0(\dr)$ is a robust
statistic, roughly independent of the spatial or velocity resolution
of the map. The power law shape of $S_0(\dr)$ extends to a larger 
physical scale, $\ell_o$, when the full size of the Blitz \& Stark 
map is used, increasing $\ell_o$ from $\approx 2.5$~pc to $\approx 10.0$~pc.
This is consistent with the correlation outer scale $\ell_o$ being
determined by the size of the map, since the Blitz \& Stark map is 
about 4 times more extended in Galactic latitude than the Heyer et al.
map. Finally, the correlation inner scale can be estimated in the
Heyer et al. map, $\ell_i\approx 0.4$~pc, while only an upper limit
can be obtained from the Blitz \& Stark map, $\ell_i< 0.7$~pc, where
the power law shape of $S_0(\dr)$ extends down to the smallest scale.
We propose that the correlation inner scale, when it is apparent, is a 
real estimate of the smallest self--similar scale in a map, and not an 
artifact of resolution. 

Figure~12 shows a significantly different spectral slope between the
Blitz \& Stark full and partial maps. The value of $\alpha$ can
in fact vary in different regions of a molecular cloud, and should
depend on various physical factors that will be discussed in detail 
in our next paper.

\nocite{Blitz+Stark86}
\nocite{Heyer+2000}

\section{Discussion}

Different statistical analyses of spectral maps, listed in \S 1, 
have been used in previous works, to 
describe quantitatively i) the hierarchical or fractal structure
of molecular clouds, and ii) their random velocity field. 
In the first case, when the spatial structure is studied, 
the velocity information is usually lost (for example by using 
maps of integrated intensity); in the second case, when the random 
velocity field is considered, information on spatial intensity 
structures is lost (for example by using only the 
velocity centroids, or the spectral shape independent of the position 
on the map). However, the dynamics of molecular clouds certainly
generate specific correlation properties in both the velocity
and density field at the same time and in a self--consistent way.
Spectral--line maps of molecular clouds can provide information
on these properties. The SCF method can offer new insight, because it 
simultaneously computes the correlation of intensity and velocity 
structure in spectral maps. The relation between the SCF method
and other statistical methods (such as wavelets and 
$\Delta$--variance) is important to understand, and is the 
subject of our next project. 

The SCF will be useful in the comparison of theoretical models with
observational data. It has already been shown in RGWW, that theoretical models
capable of generating synthetic spectra with shape (measured by their skewness
and kurtosis) similar to the shape of observed spectral
line profiles, can yield values of the SCF that are very different
from their observational counterparts. This is due to the fact that
theoretical synthetic spectra with realistic shape can be computed using
models of the density and velocity fields that are not an appropriate
description of the physical conditions in molecular clouds, and therefore
do not spatially correlate like the observational spectra.

There are a number of subtleties that must be taken into account when
the value of the SCF of synthetic spectra is computed: 
i) For ideal comparisons, noise should be added to the theoretical 
spectral--line profiles, to match the average spectrum quality ,$\aq$, 
in the observational data; ii) the spatial and spectral resolutions
in the models and in the observational counterpart should be
roughly similar, and, most importantly, iii) synthetic spectra must be 
computed for the same molecular transition that is observed.

If the instrumental noise is not uniform over any spectral
map, uniform noise should be added to the spectra
until the noise is made approximately uniform\footnote{We suggest here
rough equalization of the {\it noise} in all map pixels, not of the $S/N$ as in
RGWW.}. For the most valid comparison of observed and synthetic spectra, 
noise should be added as needed to the synthetic spectra, to make the value 
of $\aq$, equal to its value in the observational map. The dispersion around the mean 
values of the SCF are in part due to intrinsic spectral features, and in part 
to the noise. However, we have verified, by comparing with smooth 
Gaussian models, that the dominant source of the dispersion is intrinsic 
spectral features. The contribution of noise to the dispersion is small, 
as can be seen in Figure~6, since the 1--$\sigma$ ``error bars'' are almost
independent of $\aq$. In addition, Figure~6 shows that the mean value of the 
normalized SCF at $\dr =1$ is only weakly dependent on $\aq$. We have also 
shown that the slope of $S_0(\dr)$, $\alpha$, is almost independent of $\aq$
(\S 4.2). 

The SCF is a statistical tool that reflects the type of structures
present in spectral maps, and it is sensitive to the size of the structures
relative to the size (or number) of the pixels in the maps. The same region
of a molecular cloud, observed with higher resolution, can yield a higher
value of the SCF, because the neighboring spectra are closer to each
other, in physical units. In order to correctly compare theoretical models and 
observations directly, the models must be computed with a physical size
or resolution that match the observations. However, we have verified, 
in \S 4.3 and 5, that the spectral slope $\alpha$ is almost independent of 
resolution. The correct rescaling of models and observations to physical 
units of length is therefore important only for a correct estimation of 
the correlation scales $\ell_i$ and $\ell_o$, and it is not necessary for 
the determination of the value of $\alpha$.

There are yet more subtleties to be considered in applying the SCF 
to observations. Observed spectral maps are often Nyquist sampled, 
which means that beams of neighboring positions on a map overlap. 
Moreover, noise can be spatially correlated because of ``bad pixels'' 
in detector arrays, or because the data is obtained with 
``reference sharing''. Nyquist sampling, ``bad pixels'', and
``reference sharing'' increase artificially the value of the SCF
at $\dr =1$ (and possibly the determination of $\ell_i$, if this relies
on $S(\dr =1)$), but not its values for $\dr >1$. It is therefore 
possible to compute correctly $S_0(\dr >1)$ and the spectral 
slope $\alpha$, without any detailed knowledge of observational 
procedures.

We have argued in \S 4.2 and 5 that the correlation inner scale,
$l_i$, is a true estimate of the smallest self--similar scale in
a map. Different factors can determine the value of $l_i$. One 
possibility is that dense regions of size $<l_i$ are indeed
coherent (Barranco \& Goodman 1998; Goodman et al. 1998), 
in which case the value
of $l_i$ has a precise physical meaning, and may even be related
to the process of the generation of star--forming cores. Another
possibility is that the value of $l_i$ is affected by the depletion 
of the observed molecular species at high density, or by optical 
depth.    

\nocite{Barranco+Goodman98}  \nocite{Goodman+98}

\section{Summary and Conclusions}

In this work we have studied the simplest form of the SCF proposed
by RGWW ($S^0$ in RGWW), and we have discussed its dependence on
noise, velocity sampling, and spatial sampling, using observational 
data, numerical simulations of magneto--hydrodynamic turbulence, and simple 
models of Gaussian spectral line profiles. We have computed 
analytically the effect of noise on the SCF, and have proposed to 
renormalize the SCF by its analytical maximum. As a result, the SCF
now has only a weak dependence on signal--to--noise, because most 
of the effect of noise has been corrected for with the analytic 
formula. A new definition of ``spectrum quality'' has been used 
for this purpose, which we call $Q$. In the computation of the SCF 
we have never used Gaussian fits to the spectra, and we have used 
a velocity window of constant width, equal to 10 times the value 
of the standard deviation of the spectral profile averaged over 
the whole map. We have also computed the SCF as a function 
of the spatial lag between spectra, $S_0(\dr)$, which has allowed
us to describe spectral maps in terms of their spectral slope
$\alpha$, and correlation inner scale $\ell_i$.

The main conclusions of this work are the following:

\begin{itemize}
\item $S_0(\dr)$ is a power law in the range of scales $\ell_i<\ell<\ell_o$.
\item The correlation outer scale, $\ell_o$, is determined by the size
of the map; we have found no evidence for a true departure from
self--similarity on large scales, in the map analyzed so far.
\item The correlation inner scale, $\ell_i$, is a true estimate of the
smallest self--similar scale in a map.
\item The spectral slope, $\alpha$, is roughly independent of velocity
resolution, spatial resolution, and average spectrum quality, $\aq$;
it is a robust statistical property of spectral maps of molecular clouds.
\item The correlation scales, $\ell_i$ and $\ell_o$, are also roughly 
independent of velocity and spatial resolutions, and $\aq$, if $S_0(\dr)$
is properly rescaled to physical units of length.
\item Molecular transitions which trace higher gas density yield steeper
$S_0(\dr)$ power--laws than transitions tracing lower gas density.
\item Nyquist sampling, bad pixels in detector arrays, and reference 
sharing data acquisition need to be taken into account for a correct 
determination of the SCF at $\dr =1$. The value of $\alpha$, however,
can be computed correctly without a detailed knowledge of observational
procedures.
\end{itemize}

We expect that the exact value of the spectral slope and of the correlation
inner scale depend on the physical conditions in the clouds, such
as the turbulent velocity dispersion relative the the speed of sound,
the magnetic field strength, the effects of gravity, stellar outflows, 
and other influences. In our future work, the SCF will be applied to different 
molecular cloud maps and to different models, in order to study the sensitivity 
of the SCF method to physical conditions in the clouds.

\acknowledgements

We are grateful to James Moran and Ramesh Narayan for inspiring us to find
a way to ``normalize'' the SCF, and to Ur\"os Seljak for suggesting a 
formalism that led to equation 4.   We also thank J\"urgen Stutzki for 
helpful comments on this work as it was in progress. This work was 
supported by NSF grant AST-9721455.

\clearpage

\onecolumn

{\bf Figure captions:} \\

\begin{figure}
\centering
\leavevmode
\epsfxsize=1.0
\columnwidth
\epsfbox{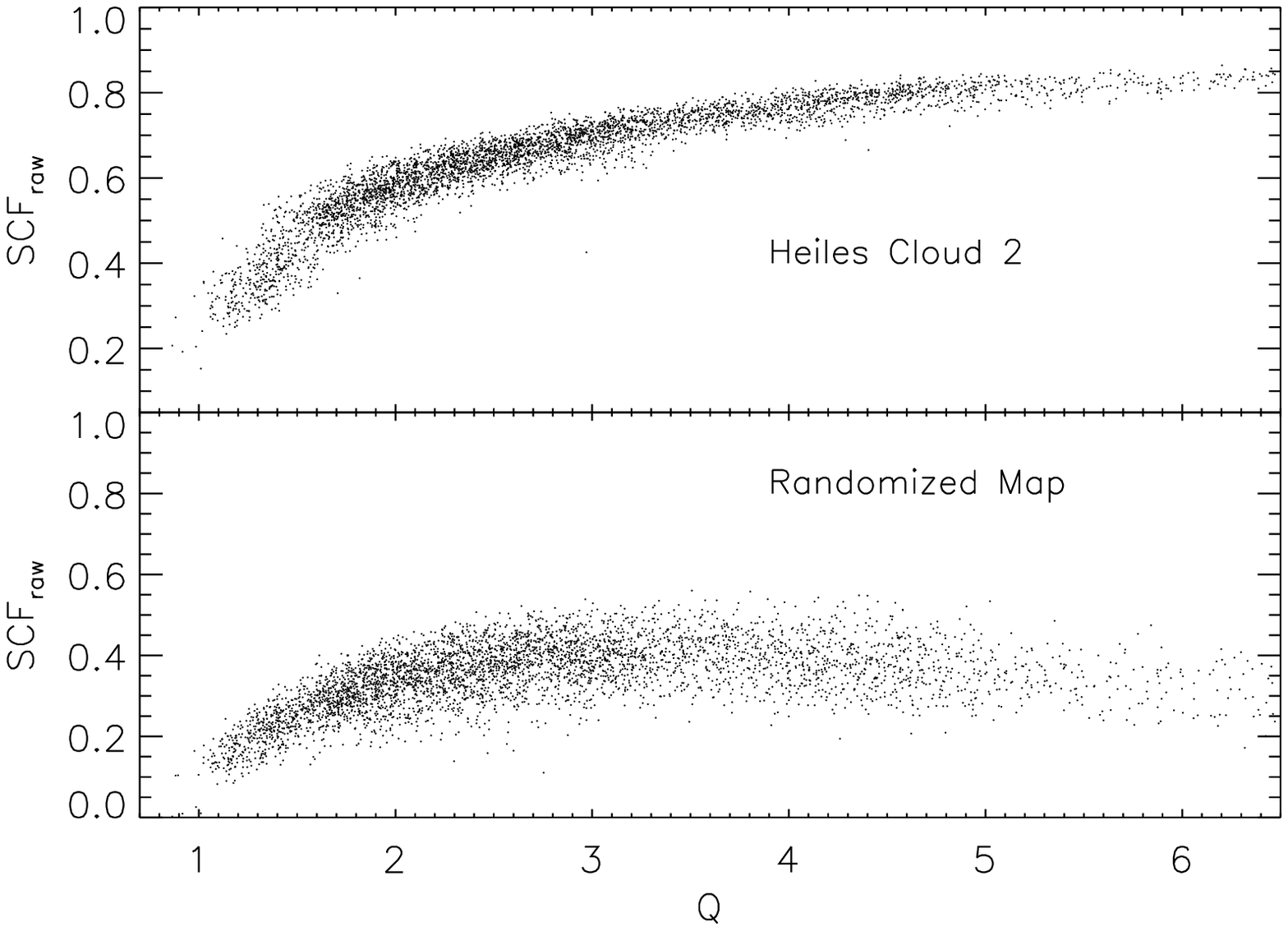}
\caption[]{}
\label{fig00}
\end{figure}

{\bf Figure \ref{fig00}:} Upper panel: SCF versus $Q$
(see the text for the definition of $Q$), for a C$^{18}$O (1-0) map of the
Heiles Cloud 2 (deVries et al. 1998). Lower Panel: The same as above, 
but with the positions of the spectra  randomized. The value of the SCF
is not normalized to its analytic approximation (see text).\\

\begin{figure}
\centerline{\epsfxsize=13cm \epsfbox{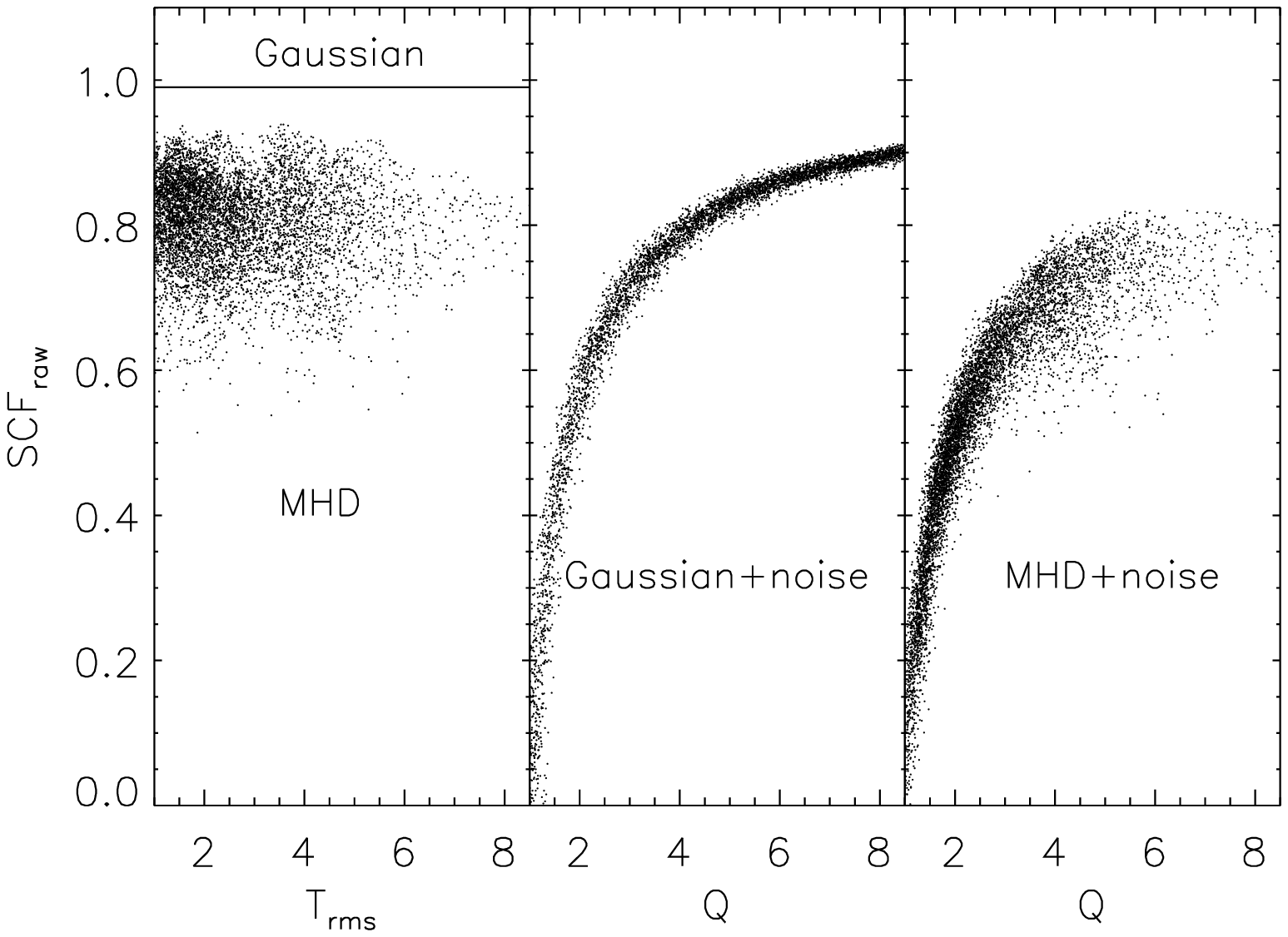}}
\caption[]{}
\label{fig6}
\end{figure}

{\bf Figure \ref{fig6}:} Left panel: SCF versus rms antenna temperature,
for a smooth map of noiseless Gaussian spectra (solid line), and for noiseless
synthetic spectra , computed using the results of
MHD turbulence simulations (scatter plot). Both the decrease of the
values of the SCF and the increase in its dispersion occurs even without
noise, and are just the effect of structure in the map (cloud structure) and
in the velocity profile of individual spectra. The map of gaussian spectra
has a linear intensity gradient along one axis, such that the total intensity
varies by a factor of 2 over the whole map.
Central and right panel:
Same as left panel, but with uniform Gaussian noise added to the spectra.
The SCF is now plotted versus $Q$.
The values of the SCF of the two models are still different
after noise has been added, which confirms that the SCF versus the $S/N$ can be
used as a tool to describe spectral line data cubes, even if the gross shape
of the scatter plots is mostly due to the effect of noise. \\

\begin{figure}
\centerline{\epsfxsize=13cm \epsfbox{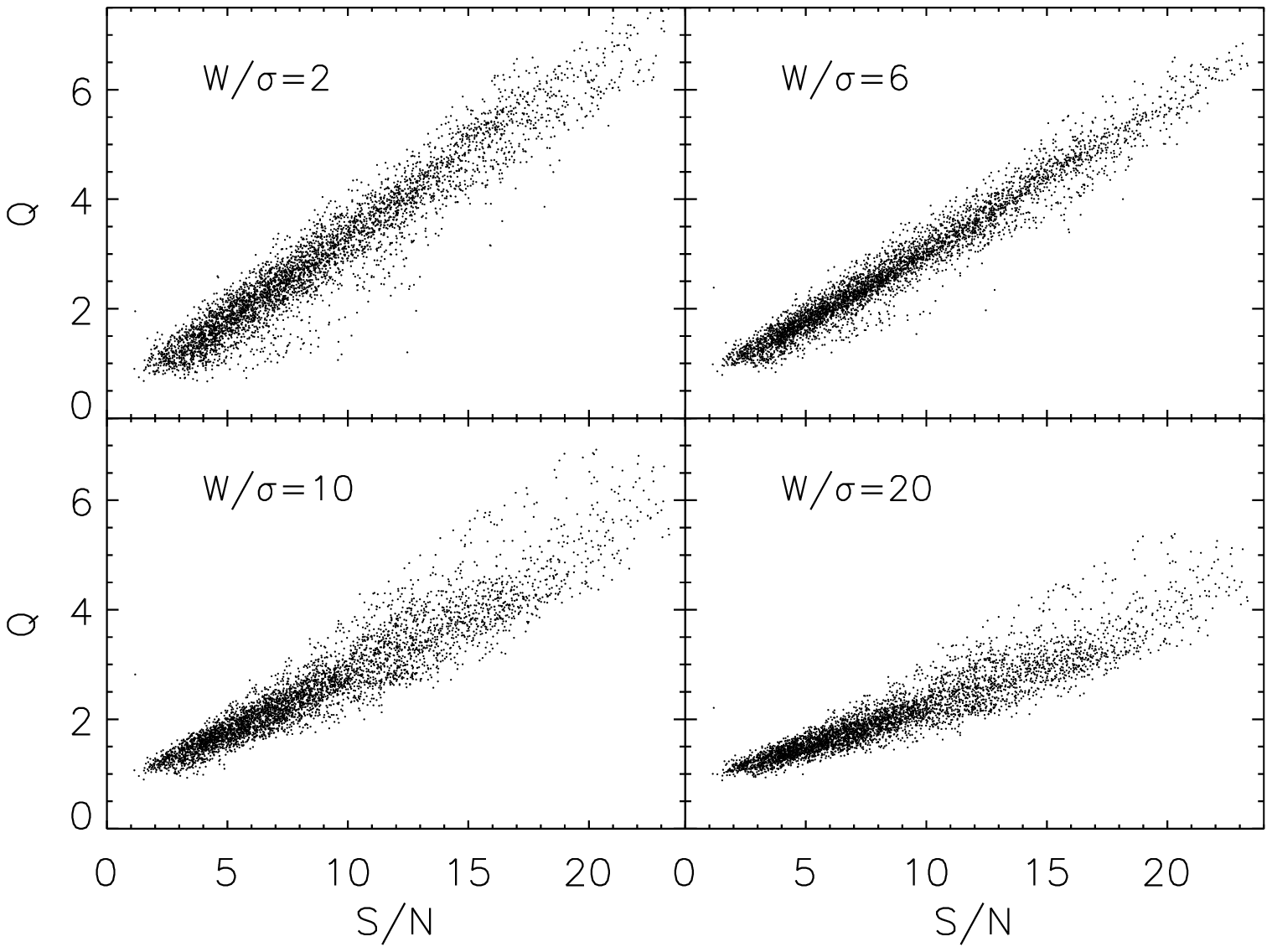}}
\caption[]{}
\label{fig2}
\end{figure}

{\bf Figure \ref{fig2}:} $S/N$, as defined in this
work, plotted against the usual definition of $S/N$,
based on Gaussian fits to the spectra. The spectra from the
Heiles Cloud 2 map are used. Each panel shows the scatter plot
for a different value of the width of the velocity window, $W$.
The parameter $\sigma$ is the standard deviation of the spectrum
averaged over the whole map. The value used to compute the SCF in
this work is $W=10\sigma$. \\

\begin{figure}
\centerline{\epsfxsize=14cm \epsfbox{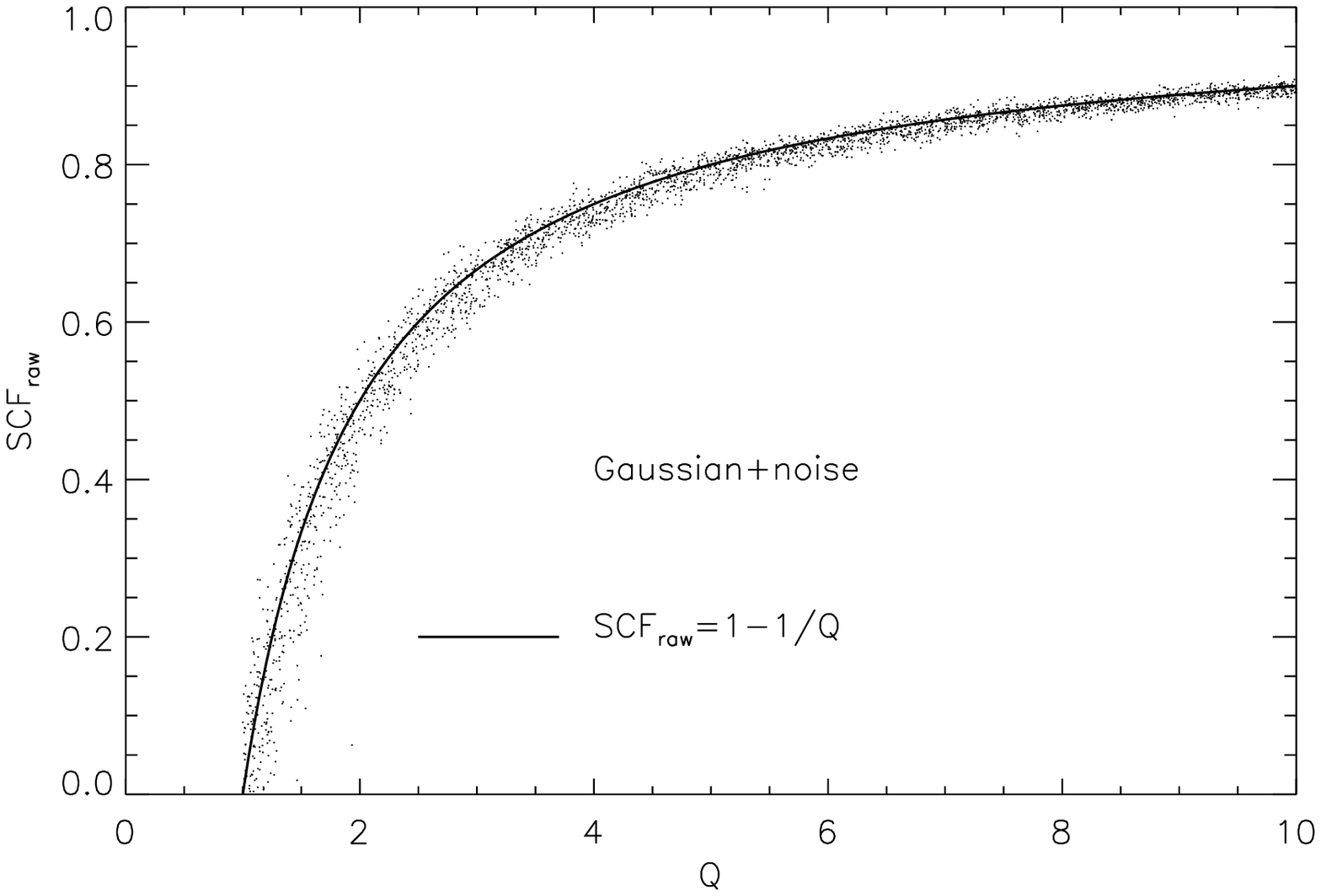}}
\caption[]{}
\label{fig10}
\end{figure}

{\bf Figure \ref{fig10}:}  SCF versus $Q$ in the model of Gaussian
velocity profiles. The continuous line is the analytic maximum SCF as a
function of $Q$,
derived under the assumption that neighboring
spectra are identical, apart from differences due only
to instrumental noise. The analytic function, $1-1/Q$, provides
an excellent fit to the Gaussian model.\\

\begin{figure}
\centerline{\epsfxsize=16cm \epsfbox{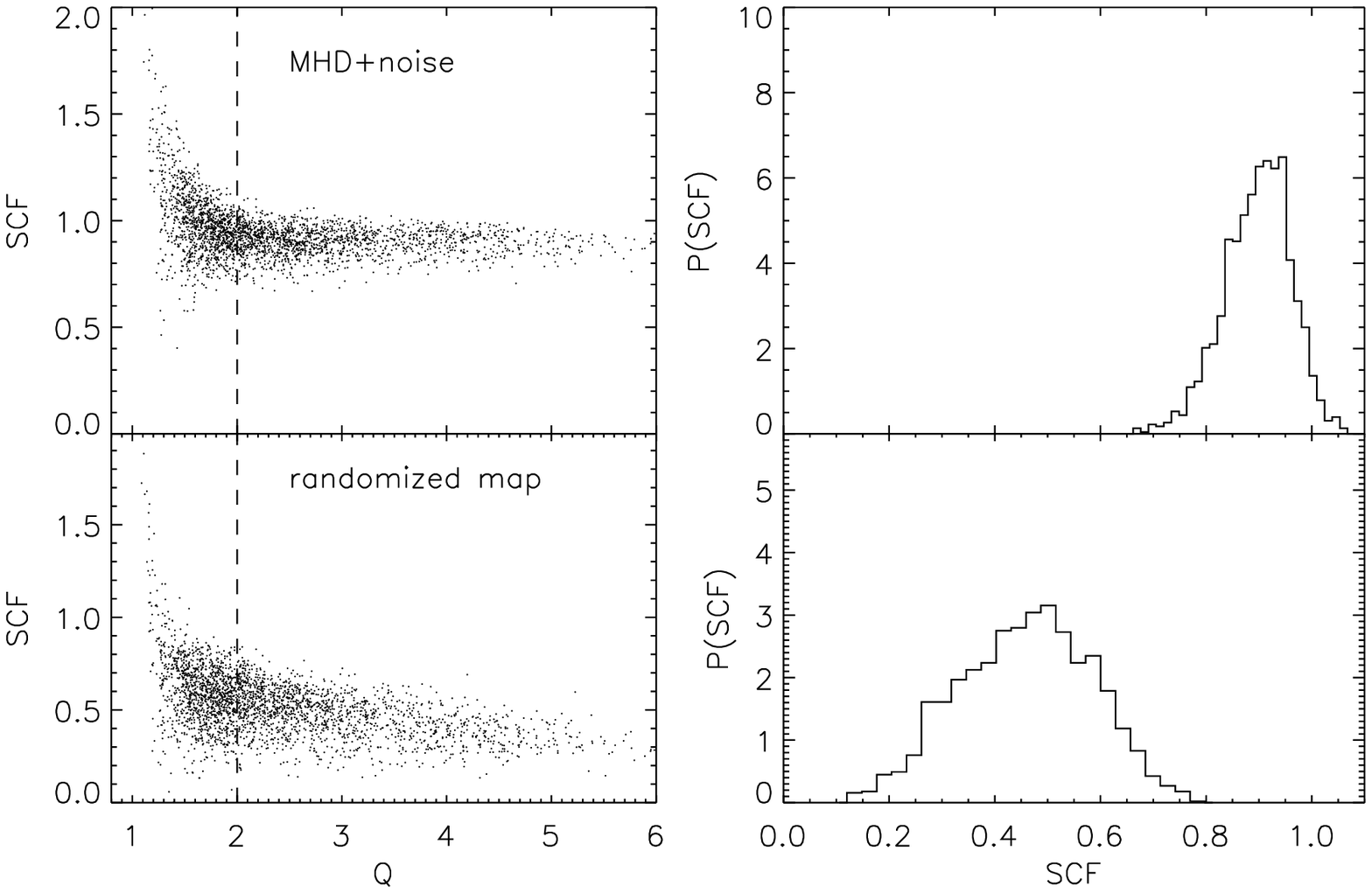}}
\centerline{\epsfxsize=16cm \epsfbox{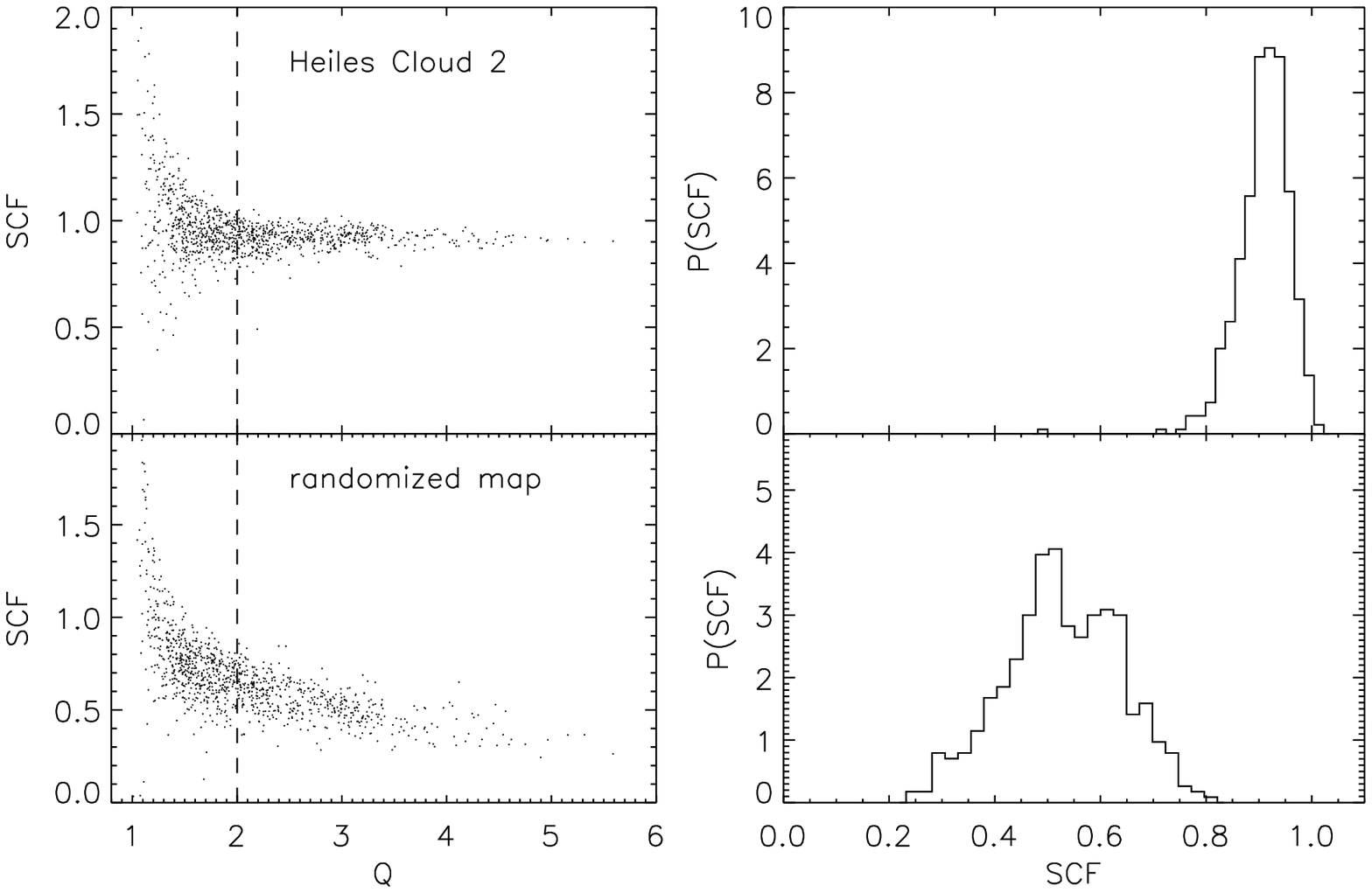}}
\caption[]{}
\label{fig9}
\end{figure}

{\bf Figure \ref{fig9}:} Left panels: {\it normalized}
SCF versus $Q$ for the MHD model (two top panels), and the Heiles
Cloud 2 data (two bottom panels). The normalized SCF is obtained
by dividing the SCF by the analytic function, $1-1/Q$.
Right panels: Histograms of the normalized SCF plotted in the
left panels. Only points from the left panels with $Q>2$ are used, 
since spectra with $Q<2$ are dominated by noise. \\

\begin{figure}
\centering
\leavevmode
\epsfxsize=1.0
\columnwidth
\epsfbox{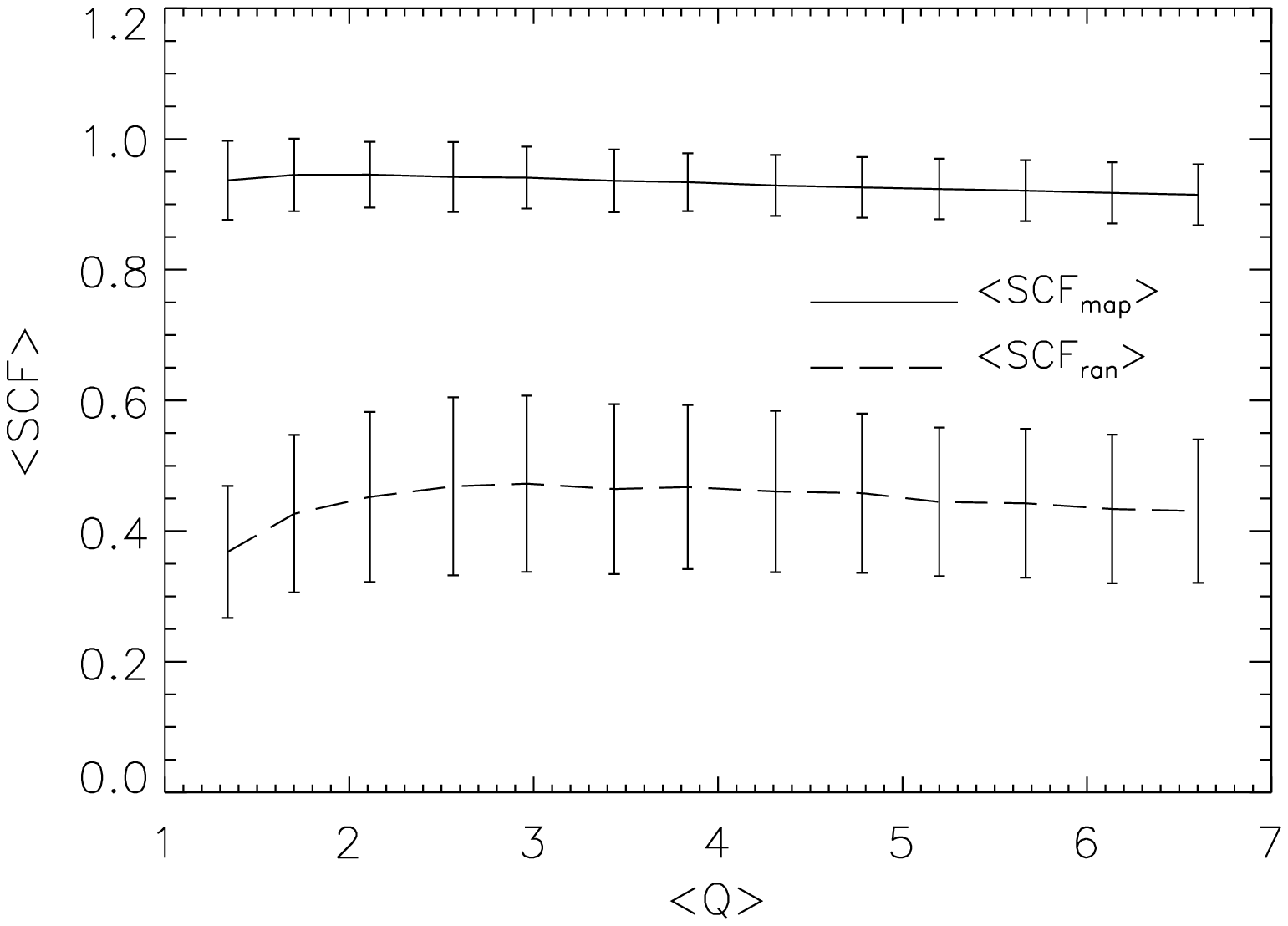}
\caption[]{}
\label{fig1}
\end{figure}

{\bf Figure \ref{fig1}:} Average values of the normalized SCF, 
using only pixels with $Q>2$, as a function
of $Q$ averaged over the whole map. The value of $\aq$ is varied by
adding different levels of noise to the synthetic spectra. The plot is
computed using a map of 90x90 synthetic $^{13}$CO spectra, from
Padoan et al. (1999).  \\

\begin{figure}
\centerline{\epsfxsize=13cm \epsfbox{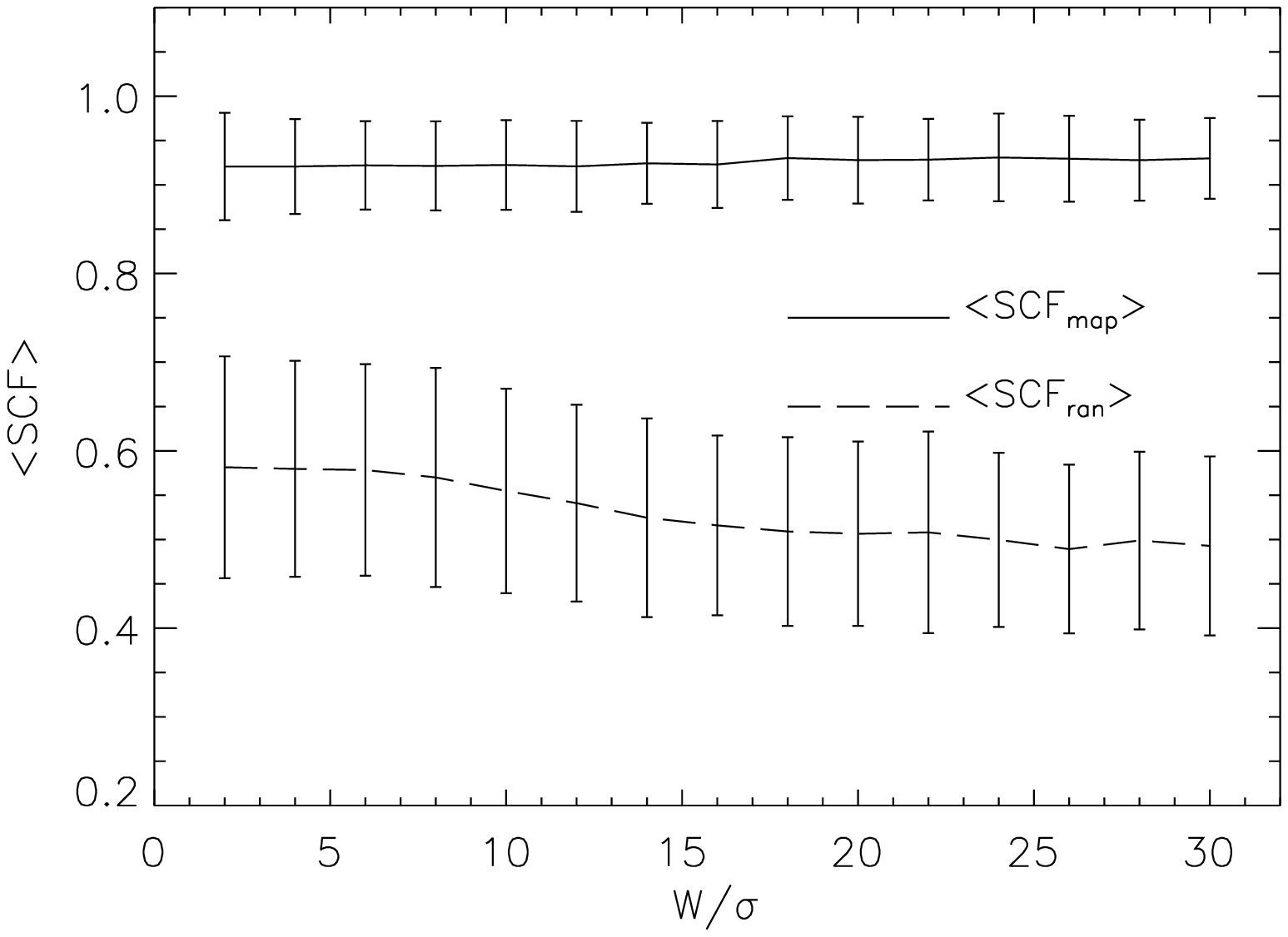}}
\caption[]{}
\label{fig3}
\end{figure}

{\bf Figure \ref{fig3}:} Average values of the normalized SCF over 
the Heiles Cloud 2 map (only spectra with $Q>2$ are used), as a 
function of the velocity window, $W$, used in the
computation of the SCF. The larger the velocity window, the smaller
the values of the SCF of the spectra with randomized positions,
because more noise enters the comparison of neighboring spectra.
This does not happen for the SCF of the original map, because the
effect of noise is appropriately corrected for.
The definition of $\sigma$ is the same as in Figure~3. \\

{\bf Figure 8:} $S_0(\dr)$ for the MHD turbulence model, the Gaussian
model, and the map of the Heiles Cloud 2, computed only for spectra 
with $Q>2$. The spectral slope of the MHD
model should not be compared with the Heiles Cloud 2 spectral slope,
although very similar, since the two maps have been obtained with different 
molecular transitions.

{\bf Figure 9:} $S_0(\dr)$ of spectral maps of different molecular 
transitions, computed only for spectra with $Q>2$.
The plot is computed using maps of 90x90 synthetic $^{12}$CO, $^{13}$CO, 
and CS spectra, from Padoan et al. (1999). The same three dimensional 
cloud model, from numerical simulations of super--sonic MHD turbulence, 
has been used in all three cases. The value of the spectral slope is 
larger for molecular transitions which trace a higher gas density than 
for transition tracing a lower density.  \\

{\bf Figure 11:} $S_0(\dr)$ for the same synthetic map, with different
average spectrum quality $\aq$. Different spectrum quality
are obtained by adding different levels of noise to the synthetic
spectra. \\

{\bf Figure 11:} $S_0(\dr)$ of spectral maps of different resolution. 
Left panel: $S_0(\dr)$ for the original Heiles Cloud 2 map (squared
symbols), and the same map rebinned to a size three times smaller
(asterisks). Right panel: Same as left panel, but for the synthetic
map.  \\

{\bf Figure 12:} $S_0(\dr)$ of the Rosette Molecular Cloud $^{13}$CO
maps by Blitz \& Stark (1986) and by Heyer et al. (2000) (see text for 
details). Triangles represent the portion of the Blitz \& Stark map
that covers the same area mapped by Heyer et al. (2000). \\

\clearpage

\clearpage

\clearpage

\clearpage

\clearpage

\clearpage

\clearpage


\clearpage
\begin{figure}
\centerline{\epsfxsize=16cm \epsfbox{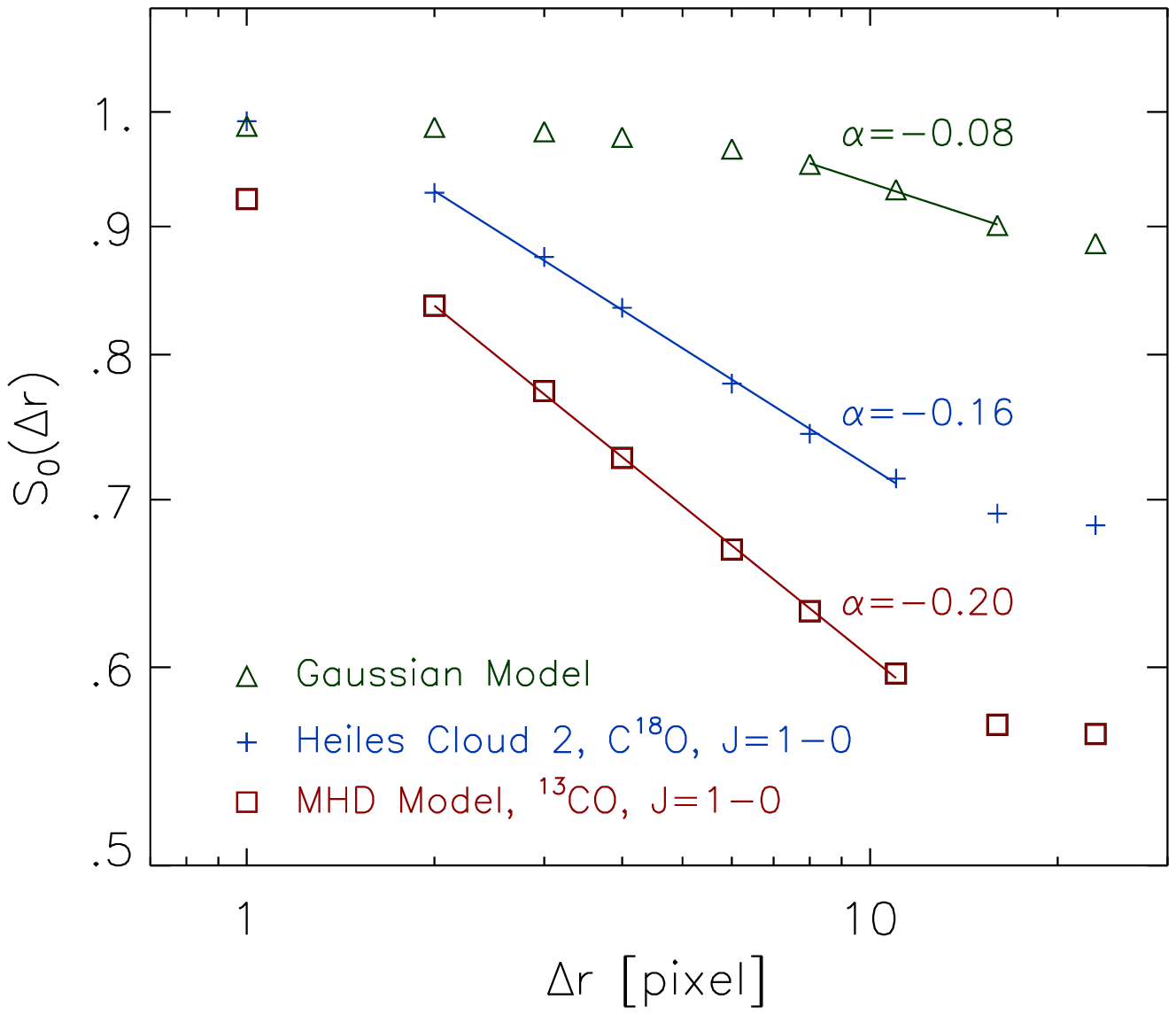}}
\caption[]{}
\label{fig30}
\end{figure}

\clearpage
\begin{figure}
\centerline{\epsfxsize=16cm \epsfbox{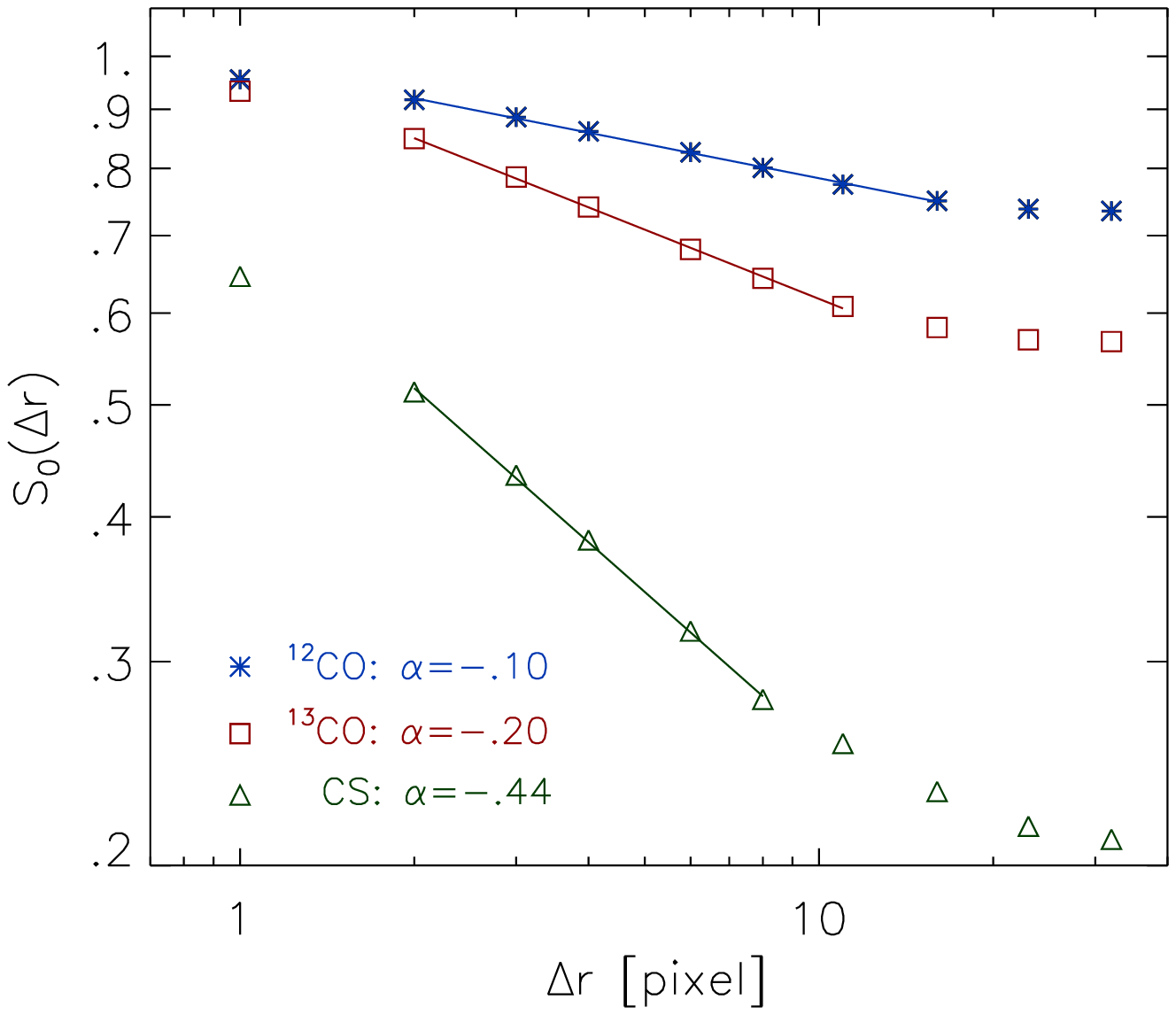}}
\caption[]{}
\label{fig32}
\end{figure}

\clearpage
\begin{figure}
\centerline{\epsfxsize=16cm \epsfbox{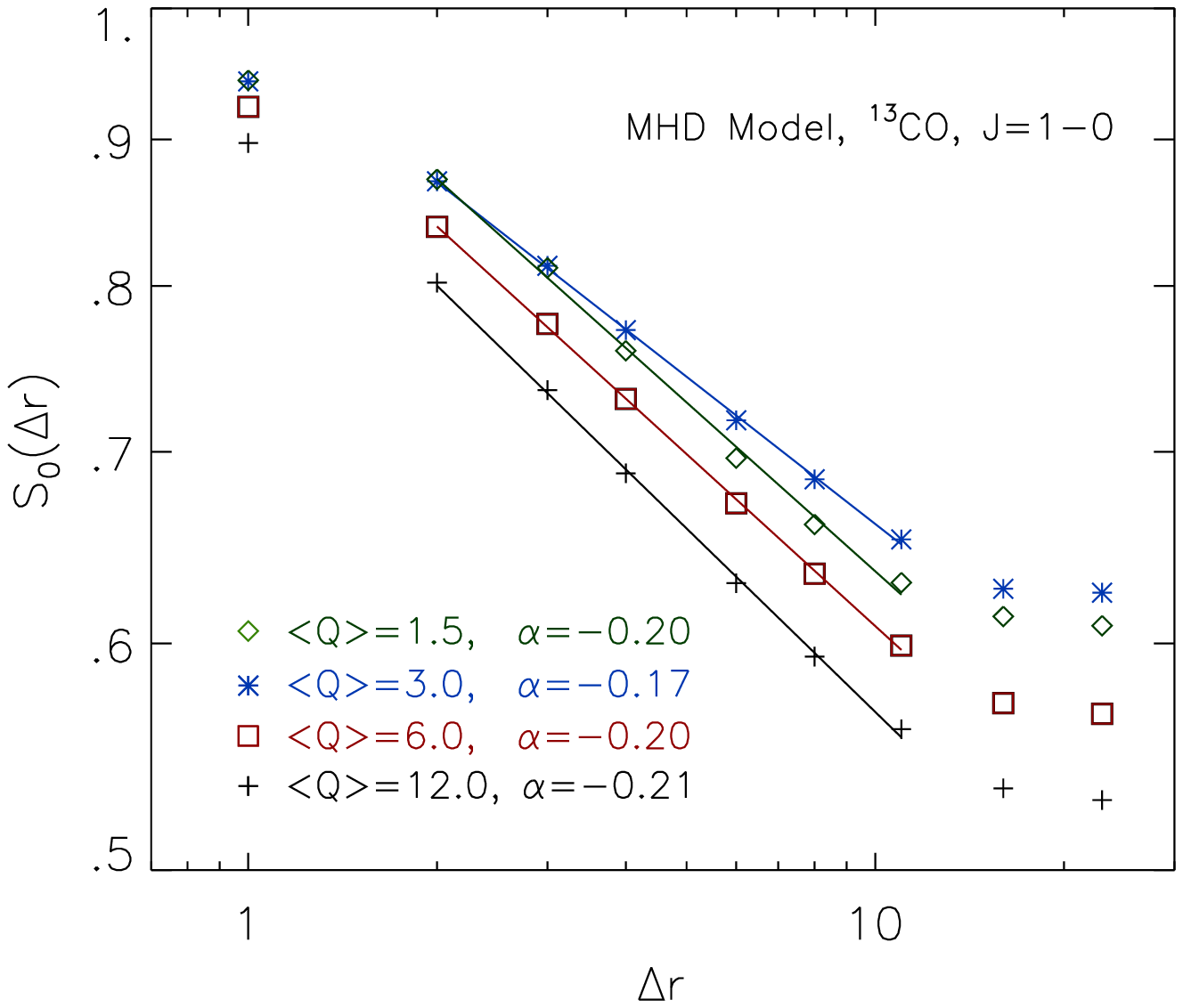}}
\caption[]{}
\label{fig32}
\end{figure}

\clearpage
\begin{figure}
\centerline{\epsfxsize=18cm \epsfbox{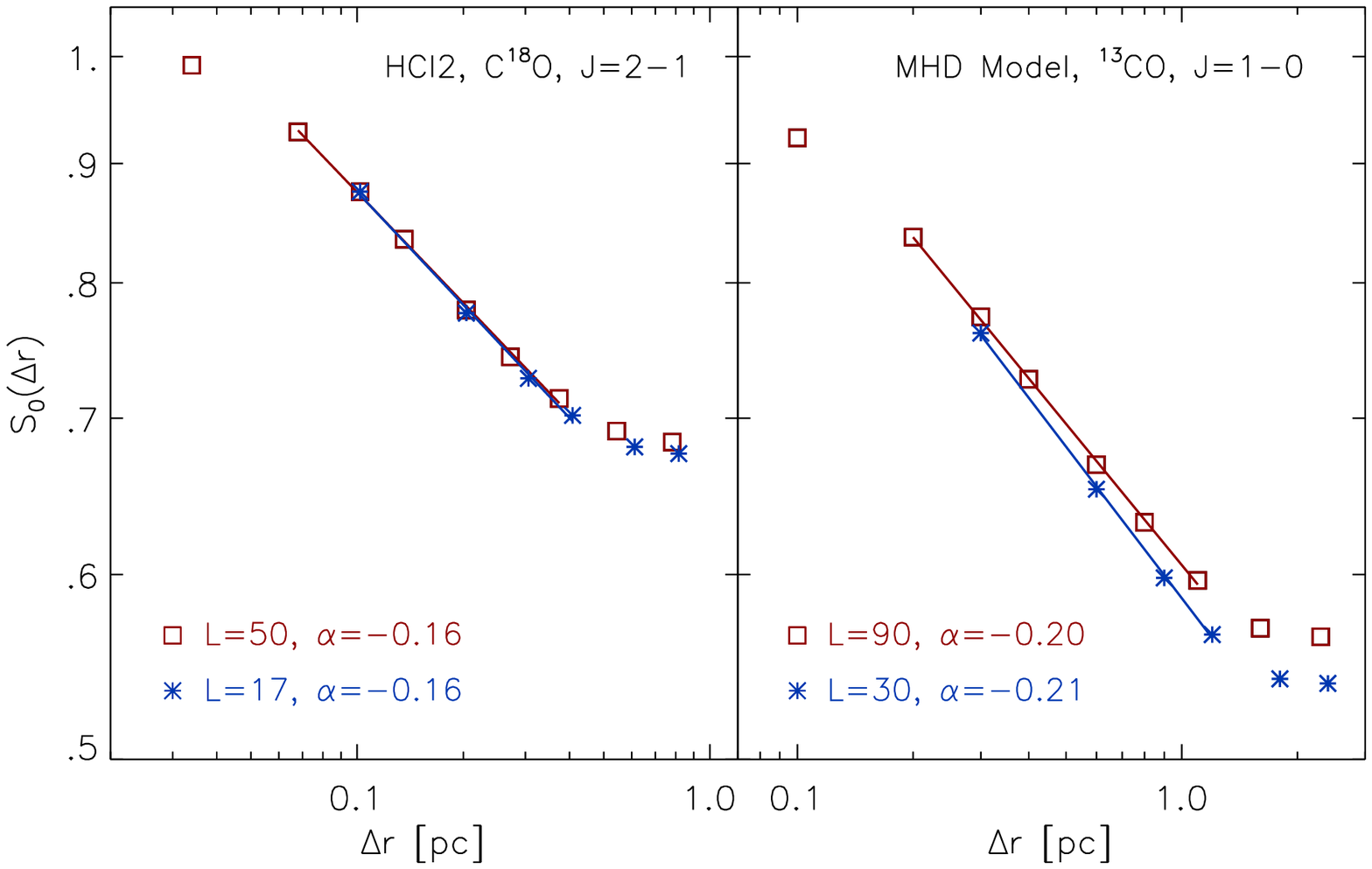}}
\caption[]{}
\label{fig31}
\end{figure}

\clearpage
\begin{figure}
\centerline{\epsfxsize=18cm \epsfbox{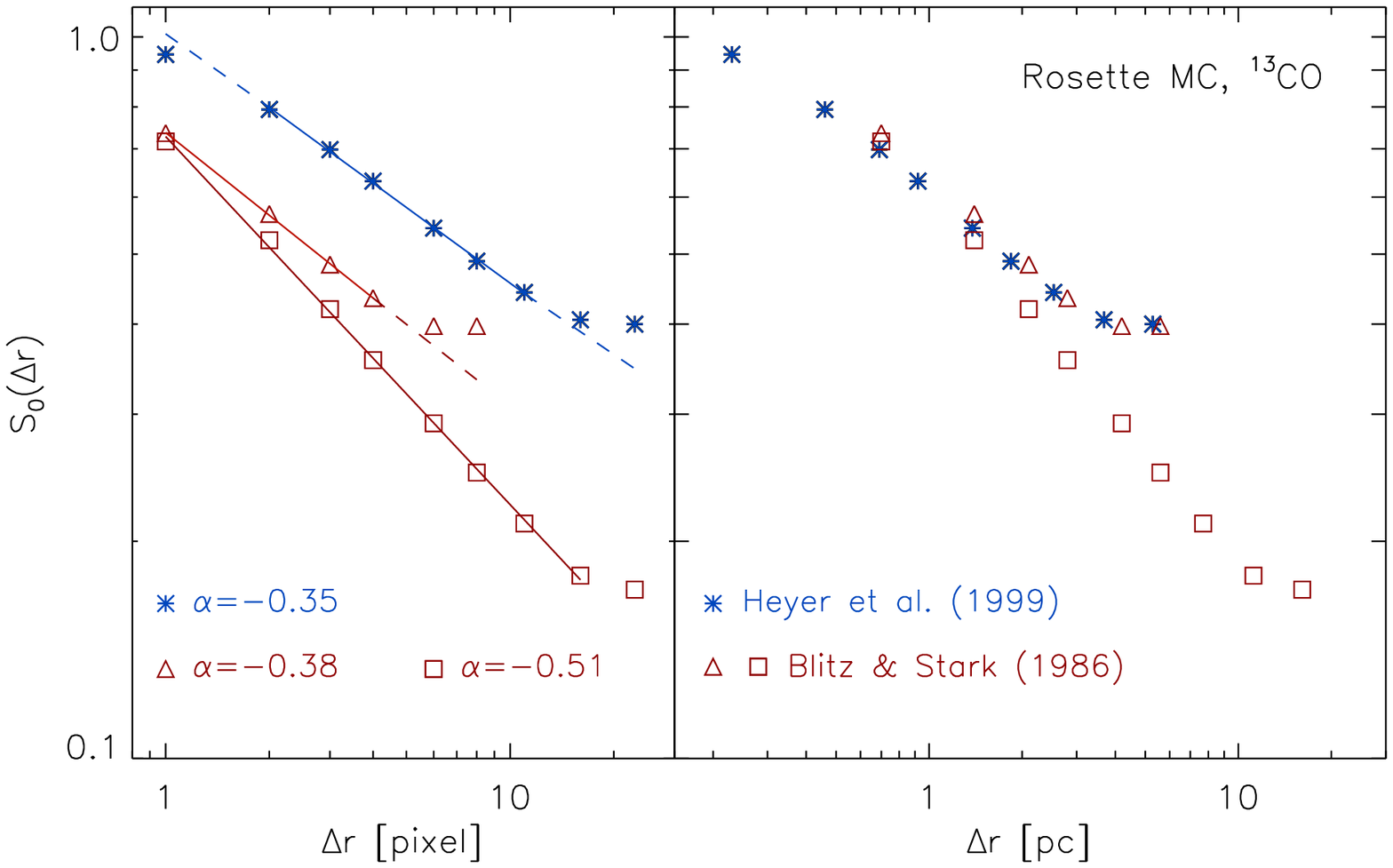}}
\caption[]{}
\label{fig33}
\end{figure}

\end{document}